\documentclass[twocolumn,showpacs,aps,pre,nofootinbib]{revtex4}
\newcommand{\al}{\alpha}
\newcommand{\be}{\begin{equation}}
\newcommand{\ee}{\end{equation}}
\newcommand{\la}{\lambda}
\newcommand{\del}{\delta}
\newcommand{\om}{\omega}
\newcommand{\ep}{\epsilon}
\newcommand{\pd}{\partial}
\newcommand{\bra}{\langle}
\newcommand{\ket}{\rangle}

\newcommand{\non}{\nonumber}
\newcommand{\bea}{\begin{eqnarray}}
\newcommand{\eea}{\end{eqnarray}}

\begin{document}

\title{Fermi-Pasta-Ulam $\beta$ lattice: Peierls equation and 
anomalous heat conductivity}
\author{Andrey Pereverzev}
\email{andrey.pereverzev@trinity.edu}
\affiliation{Center for Studies in Statistical Mechanics and Complex 
Systems, University of Texas at Austin, Austin, Texas 78712}
\affiliation{Department of Chemistry,
Trinity University, San Antonio, Texas 78212}

\date{\today}

\begin{abstract}

The Peierls equation is considered for the Fermi-Pasta-Ulam $\beta$ 
lattice. Explicit form of the linearized collision operator is obtained.
Using this form the decay rate of the normal mode energy as 
a function of wave vector $k$ is estimated to be proportional to $k^{5/3}$. 
This leads to the  $t^{-3/5}$ long time behavior of the current correlation 
function, and, therefore, to the divergent coefficient of heat conductivity. 
These results are in good agreement with the results of recent computer 
simulations. Compared to the results obtained through 
the mode coupling theory our estimations give the same $k$ dependence 
of the decay rate but a different temperature dependence.
Using our estimations 
we argue that adding a harmonic
on-site potential to the Fermi-Pasta-Ulam $\beta$ lattice may lead to 
 finite heat conductivity in this model.
\end{abstract}
\maketitle

\section{Introduction} \label{intro}
The Peierls equation has played an important role in understanding 
properties of
solids  since its original derivation by Peierls \cite{Peierls1,Peierls2}. 
It was successfully used for  qualitative explanation  of heat conduction
 in dielectrics and for prediction of  such phenomena as second sound and 
Poiseuille flow
\cite{Beck}. In spite of these successes quantitative predictions are hard 
to make due to the enormous complexity of the equation even for solids with
 simple 
dispersion laws. It is well known that approximating  the solid by isotropic 
continuum 
leads to divergent heat conductivity even in three dimensions
if only three phonon collisions are considered \cite{Peierls2,Lifshits}. 
This divergence
can be eliminated if more careful analysis of the dispersion relations is 
performed \cite{Herring}.
In this paper we would like to consider the collision 
operator of the linearized
Peierls equation for a simple one dimensional model: linear chain with 
 quartic 
interaction also known as Fermi-Pasta-Ulam (FPU) $\beta$ lattice.

One dimensional lattices has drawn considerable attention since the original
work of  Fermi, Pasta, and Ulam \cite{FPU} and originated a vast area 
of research \cite{Ford}. It is now well established by computer 
simulations that the heat conductivity
in FPU lattices diverges when the size of the lattice goes to infinity
\cite{Lepri1,Lepri3,Lepri4,Lepri5,Aoki}. The simulations give a
power law dependence of the heat conductivity on the number of
particles $N$ as approximately $N^{2/5}$. This form of $N$ dependence is
related to the $t^{-3/5}$ long time behavior of the current correlation 
function. Theoretical work in this area \cite{Lepri3,Lepri4} was based 
on the application of the mode coupling theory \cite{Balescu1,Resibois1}.
Application of Peierls equation to the analysis of heat conduction 
in FPU lattices was limited to qualitative estimations 
\cite{Peierls1,Lepri4,Jackson}. 
It is therefore of interest to check if more careful analysis 
of the Peierls equation can explain some of the anomalous properties 
of the FPU chains.

To this end  we will consider the explicit form of the Peierls
equation for the FPU lattice in Sec. II. In Sec. III we apply this 
equation to estimate the long time behavior of the current correlation 
function as well as the wave vector dependence of the decay rate for mode 
energies. Concluding remarks are given in Sec. IV.

\section{The Peierls equation}
The Hamiltonian for the FPU $\beta$ lattice is 
\be
H=
\sum_{r}\frac{{p_r}^2}{2m}+
\sum_{r}\frac{C}{2}\left(u_{r+1}-u_{r}\right)^2
+\sum_r\frac{\la}{4}\left(u_{r+1}-u_{r}\right)^4.
\label{Ham1}
\ee
Here $u_r$ is the displacement of the particle at site $r$, $ p_r$ is the 
momentum conjugate
to
$u_r$ and
$m$ is the mass of the particle.  We also use $\la$ as a coupling
constant.
Cyclic boundary conditions are imposed, i.e.,
$u_{r+N}=u_{r}$,
where $N$ is the number of particles.
We can introduce action  variables $J_{k}$ and angle variables $\al_{k}$ 
related to $u_r$ and $p_r$ through
\bea
\sum_r u_{r}e^{-i{kr}}
&=&\sqrt{\frac{N}{2m}}
\left(\sqrt{\frac{J_{k}}{\om_{k}}}e^{i\al_{k}}
 +\sqrt{\frac{J_{-k}}{\om_{
-k}}}e^{-i\al_{- k}}\right), \non \\ 
\sum_r p_re^{i{kr}}&=&
i\sqrt{\frac{mN}{2}} \non \\
& &\times\left(\sqrt{ J_{k}\om_{
k}}e^{-i\al_{k}}
 - \sqrt{J_{-k}\om_{
-k}}e^{i\al_{- k}}\right).\non \\
& &
\eea
Here
$
k={2\pi n}/{N} \label{wave vector}
$
is a dimensionless wave vector and $n$ is an integer. The wave vector
 is usually restricted to the interval
$-\pi< k \leq \pi$ but any other interval of length $2\pi$ can be 
chosen.
The frequencies $\om_k$ are given by
\be
\om_k=
2\sqrt{\frac{C}{m}}\left|\sin\frac{k}{2}\right|. \label{frec}
\ee
In the action and angle variables Hamiltonian (\ref{Ham1}) has the form
\begin{widetext}
\be
H =\sum_{k}\om_{k}J_{k}
+\ \la\!\!\!\!\!\!\sum_{\stackrel{kk'k''k'''}{\ep \ep' \ep''\ep''' = \pm1}}
\!\!\!\!\!\!V_{\ep k\ep'k'\ep''k''\ep'''k'''}
\sqrt{\frac{J_{k}J_{k'}J_{k''}J_{k'''}}
{\om_{k}\om_{k'}\om_{k''}\om_{k'''}}}
e^{i(\ep\al_{k}+\ep'\al_{k'}+\ep''\al_{k''}+\ep'''\al_{k'''})} 
 \label{Ham2}
\ee
Coefficient $V_{ \ep k\ep' k'\ep'' k''\ep'''k'''}$ is given by
\bea
V_{\ep k\ep' k'\ep'' k''\ep'''k'''}&=&
\frac{1}{Nm^2}\sin\frac{\ep k}{2}\sin\frac{\ep'k'}{2}
\sin\frac{\ep''k''}{2}\sin\frac{\ep''' k'''}{2} 
\cos\frac{ \ep k+\ep' k'+\ep'' k''+\ep'''k'''}{2} \non \\
& &\times\Delta_{\ep k+\ep'k'+\ep''k''+\ep'''k'''}. \label{potential}
\eea
\end{widetext}
Here $\Delta_{l}$ is given in terms of Kronecker deltas as 
\be
\Delta_{l}=\sum_m\delta_{2\pi m,l}
\ee
with $m$ being an integer. Note that only terms with  $m=0$ and $m=\pm1$ 
have to be considered in (\ref{potential}). Indeed, the maximum length for 
the sum of four wave vectors is $4\pi$ but coefficient (\ref{potential}) 
vanishes in this case. 

In the problem of heat conduction the quantities of interest are the 
mode energy $E_k=\om_kJ_k$ and the total heat current given by
\be
j_h=\sum_kv_k\om_kJ_k=\sum_kv_kE_k,\label{harcur}
\ee
where $v_k$ is the group velocity.
Note that $E_k$ and $j_h$ represent only the harmonic parts of 
the corresponding quantities, it is assumed that the contributions from 
the anharmonic corrections are small for weak coupling. 
The approximate time evolution of the average energy of 
the normal mode for weak coupling and for the  lattice with no temperature
gradient and  close to the thermal 
equilibrium is given by the homogeneous linearized Peierls 
equation \cite{Peierls1,Peierls2,Lifshits}. 
The equation  is usually considered in the context of quantum mechanics
for lattices with cubic anharmonicity.
Derivation of this equation for the lattice with classical 
Hamiltonian (\ref{Ham2})
is straightforward. Note that in this case  the following 
conditions on wave vectors and frequencies have to be satisfied 
simultaneously
\bea \pm k\pm k'\pm k''\pm k'''&=&0 
\mbox{, or} \pm 2\pi,  \non \\
\pm \om \pm \om'\pm \om''\pm \om''&=&0 \label{encon}
\eea
 with the same ordering of 
signs for both relations. With the $k$ dependence of 
frequencies (\ref{frec}) the relations can be satisfied only 
when two plus signs  and two minus signs appear in  (\ref{encon}),
 i.e., in quantum mechanical terms, only the processes 
conserving the number of phonons contribute. In addition, although normal
processes exist for this number conserving case, they only result in 
exchange of quasi momenta between two colliding phonons, and, therefore,
cannot change the phonon distribution. Thus, only umklapp number conserving 
processes contribute in the collision integral of the Peierls equation. 
We can write the linearized Peierls equation for the average 
energy $\overline{E}_k$ of mode $k$, where overlining denotes averaging
over a distribution function. The equation has the form  
\begin{widetext}
\bea
\frac{\pd \overline{E}_k}{\pd t}= 
-\frac{{(24\la k_BT)}^2\pi}{N^2m^4} 
\sum_{k'k''k'''}\Biggl[\!\!\!\!\!\!& &\frac{|V_{kk'-k''-k'''}|^2
\del(\om_k+\om_{k'}-\om_{k''}-\om_{k'''})}
{\om_k\om_{k'}^2\om_{k''}^2\om_{k'''}^2} \non \\
& &\times\left(\om_k\overline{E}_k+\om_{k'}\overline{E}_{k'}
-\om_{k''}\overline{E}_{k''}-\om_{k'''}\overline{E}_{k'''}\right) 
\Biggr].\label{pe}
\eea
\end{widetext}
The collision operator on the right hand side of equation (\ref{pe})
is a Hermitian operator in the Hilbert space with the inner product
given by
\be
\bra g|f\ket=\int_{-\pi}^{\pi}\!dk\,g^*_kf_k.
\ee
It can be shown that the collision operator has a continuous spectrum
that is bounded form below by zero \cite{Buot,Jackle}.
We wrote equation (\ref{pe}) in the form that makes it easy to see 
that $\overline{E}_k=const$ and $\overline{E}_k=const/\om_k$ are  
eigenstates of the collision 
operator with zero eigenvalues. The first eigenstate corresponds to 
the conservation
of the total energy. The second one corresponds to the conservation of  
the sum of action variables for all modes (or, in quantum mechanical 
language, to the conservation of the number of phonons). Note that the 
second eigenstate has an infinite norm.

We can write the average energy as 
\be
\overline{E}_k=k_B T+\delta \overline{E}_k, \label{deviation}
\ee
where $k_B T$ is the equilibrium value of $\overline{E}_k$ and 
$\delta \overline{E}_k$ is a deviation from that value.
If we want the average energies to approach their equilibrium value of
$k_B T$ for long times then  $\delta \overline{E}_k$ should be orthogonal 
to both of the zero eigenvalue eigenstates of the collision 
operator \cite{Lifshits}, i.e., we must have
\be
\int_{-\pi}^{\pi}\!dk\,\delta \overline{E}_k=0,\qquad
\int_{-\pi}^{\pi}\!dk\,\frac{\delta \overline{E}_k}{\om_k}=0.  
\ee

The delta function appearing in (\ref{pe}) is meaningful only in the limit 
of $N\to\infty$.
In this limit we replace the sums by integrals and Kronecker deltas by
delta function 
according to
\be
\frac{2\pi}{N}\sum_k \to \int\! dk, \qquad 
\frac{N}{2\pi}\del_{k,k'}\to\delta(k-k').
\ee
After the limit is taken we will have  terms containing 
products of two delta functions in the integrand. Therefore, 
two integration can be rather easily performed.
Integrations can be done much easier and the resulting expressions have
a simpler form if the wave vector is restricted to the interval
from $0$ and $2\pi$ rather then from $-\pi$ to $\pi$. Using explicit 
expressions for $V_{kk'-k''-k'''}$ and $\om_k$ we obtain after tedious 
but straightforward calculations 
\begin{widetext}
\bea
\frac{\pd \overline{E}_k}{\pd t}=-\frac{(3\la k_BT)^2\sin\frac{k}{2}}
{4\pi C^{7/2}m^{1/2}} 
\!\!\!\!& &\left(\int _{int}\!dk'
\frac{\sin\frac{k'}{2} \overline{E}_{k'}}
{\sqrt{\frac{1}{4}\left(\cos\frac{k}{2}+\cos\frac{k'}{2}\right)^2
-\sin\frac{k}{2}\sin\frac{k'}{2}}}\right. \non \\
& &\,\,-\left.\int_0^{2\pi}\!dk'
\frac{\sin\frac{k'}{2} \overline{E}_{k'}
-\frac{1}{2}{\sin}\frac{k}{2} \overline{E}_k}
{\sqrt{\frac{1}{4}\left(\cos\frac{k}{2}+\cos\frac{k'}{2}\right)^2
+\sin\frac{k}{2}\sin\frac{k'}{2}}}\right).\label {pco}
\eea
\end{widetext}
Subscript $int$ in the  first integral in (\ref{pco}) means that the
integral is taken over the interval where the 
integrand is real. This interval consists of two segments: from $0$ to 
$l_{1}(k)$ and from  $l_{2}(k)$
to $2\pi$, where $l_{1}(k)$ and  $l_{2}(k)$ are the two solutions of the
transcendental equation for $k'$
\be
\frac{1}{4}\left(\cos\frac{k}{2}+\cos\frac{k'}{2}\right)^2
-\sin\frac{k}{2}\sin\frac{k'}{2}=0.
\ee
 The solutions, which depend on $k$ as a parameter, satisfy 
$l_{1}(k)\le l_{2}(k)$. In the next section we will use 
equation (\ref{pco}) to estimate the long time behavior of the heat
current correlation function and the $N$ dependence of the coefficient
of thermal conductivity.

\section{The long time behavior of the correlation function}
The coefficient of thermal conductivity can be calculated by  using the 
current correlation function.
The correlation function is defined as
\be
D_N(t)=\frac{1}{N}\int\! d{\{J_k\}}d{\{\al_k\}}\,j(t)j(0)
\frac{e^{-\frac{H}{k_B T}}}{Z}.
\ee
Here $\{J_k\}$ and $\{\al_k\}$ denote the set of action and angle 
variables for all the modes, $Z$ is the partition function  for the 
equilibrium ensemble and $j$ is the total energy current. 
The coefficient of heat conductivity is given by
\be
\kappa=\frac{1}{k_B T^2}\lim_{t\to\infty}\int_0^t\!d\tau\! 
\lim_{N\to\infty}D_N(\tau). \label{hcc}
\ee
We can rewrite $D_N(t)$ as \cite{Resibois}
\be
D_N(t)=\frac{1}{N}\int\! d\{J_k\}d\{\al_k\}\,\xi j(0)
\tilde\rho(t) \label{corfu}
\ee
with
\be
\tilde\rho(t) = e^{-iLt}
\left(\frac{j(0)}{\xi}\frac{e^{-\frac{H}{k_B T}}}{Z}
+\frac{e^{-\frac{H}{k_B T}}}{Z}\right), \label {disfu}
\ee
where $L$ is the Liouville operator corresponding to Hamiltonian 
(\ref{Ham2}) and $\xi$ is an auxiliary parameter insuring the 
correct dimensions for $\tilde\rho(t)$. The parameter does not
appear in the final expressions. In going from (\ref{hcc})
to (\ref{corfu}) we also used the fact that the average current 
over the equilibrium distribution is zero.
Equation  (\ref {corfu}) shows that the correlation function can be
expressed through the average  
current per particle with the averaging performed 
 over the nonequilibrium distribution function (\ref{disfu}).
If we approximate the total heat current by its harmonic part  (\ref{harcur})
we can see that it depends only on 
the action variables and, therefore, the time evolution of the correlation 
function can be reduced to the time evolution of the average mode energies 
which is governed by (\ref{pco}). Note that if only harmonic terms are kept 
in distribution function (\ref{disfu}) at $t=0$ then the initial average
energy for mode $k$ is
\be
\overline{E}_k(0)=k_BT+\frac{2 v_kk^2_BT^2}{\xi}.
\ee 
This has the form given in (\ref{deviation})
with the deviation from $k_BT$ orthogonal to both of the zero eigenvalue  
eigenstates of 
the collision operator. Therefore, we expect the average mode energies
to approach 
$k_BT$ for long times.
To estimate the time behavior of $\overline{E}_k$ based on (\ref{pco}) 
we will use the relaxation time approximation \cite{Beck}. 
We assume that the energy of each mode approaches zero with a 
characteristic time
$\tau_k$ which depends on the
wave vector, i.e.,
\be
\frac{\pd \overline{E}_k}{\pd t}\approx
-\frac{1}{\tau_k}(\overline{E}_k-k_BT).
\ee
Some plausibility arguments in support of this approximation were
given in \cite{Klemens,Herring}.
In this approximation and for $N\to\infty$  the correlation function 
(for which we now drop the subscript $N$) is given by 
\be
D(t)=\frac{2k^2_BT^2}{\pi}
\int_0^{\pi}\!dk\,e^{-\frac{t}{\tau_k}}v_k^2. \label{intass}
\ee
Since the decay rate for energy of the normal mode with $k=0$ is zero 
(due to the conservation of the total momentum) we can expect $1/\tau_k$ 
to behave as some positive
 power of $k$ for small $k$. The long time behavior of $D(t)$ 
in (\ref{intass}) will be determined by the small $k$ behavior of $1/\tau_k$.
 Following reference \cite{Herring} we will further
assume that the $k$ dependence of $1/\tau_k$
for small $k$ is the same as in the multiplicative part of the collision
operator in (\ref{pco}).

Both the relaxation time approximation and the assumption that
the $k$ dependence of the relaxation rate for small $k$  is the same as
in the multiplicative part of the collision operator has been 
widely used in the theory of heat conduction in insulators 
\cite{Beck,Herring}. A convincing
justification of both assumptions , however, is lacking. 
Reference \cite{Herring} tries to justify both assumptions at least for
wave vectors with small $k$ by the following  reasoning. 
If only the multiplicative part was kept in the collision operator the 
resulting equation would describe a physical situation when all modes 
except mode $k$ are in equilibrium. In general this is 
not the case. However, for any initial nonequilibrium distribution of 
energy all modes except those with very small $k$ quickly relax to 
equilibrium. As a result as far as the small $k$ modes  are
concerned after a short time the physical situation is similar to the 
one just described and
 the integral 
part of the
collision operator becomes negligible compared to the multiplicative part.

If we accept both approximation for equation (\ref{pco}) then we expect  
$1/\tau_k\propto \sin^2(k/2)I(k)$ with
\be
I(k)=\int_0^{2\pi}dk'\!
\frac{1}
{\sqrt{\frac{1}{4}\left(\cos\frac{k}{2}+\cos\frac{k'}{2}\right)^2
+\sin\frac{k}{2}\sin\frac{k'}{2}}}.
\ee
This integral can be reduced 
to an elliptic integral of the first kind through 
the substitution $x=\tan(k'/4)$,
\be
I(k)\propto\int_0^{\infty}\!dx\frac{1}{\sqrt{P_k(x)}}, \label{elliptic}
\ee
where
\bea 
P_k(x)\!&=&\!
\left(1-\cos\frac{k}{2}\right)^2\!x^4+8\left(\sin\frac{k}{2}\right)x^3 \non \\
& &-\,2\!\left(\sin^2\frac{k}{2}\right)x^2
+8\left(\sin\frac{k}{2}\right)x+\left(1+\cos\frac{k}{2}\right)^2\!\!.\non \\
& &\label{polynomial}
\eea
Integral (\ref{elliptic}) can be reduced to the Legendre
normal form and its $k$ dependence can be expressed in terms of
 the $k$ dependence for the roots of the forth order polynomial
(\ref{polynomial}) \cite{Byrd}.
Since the calculations are rather
involved and we are interested only in  the small $k$ behavior
of $I(k)$ we give here
a less rigorous but simpler estimation that gives the same result for
small $k$. We expand the coefficients in the polynomial in
powers of $k$ and keep the lowest order terms in front of each monomial to
get
\be
I(k)\propto\int_0^{\infty}\!dx\frac{1}
{\sqrt{\frac{k^4}{64}x^4+4kx^3-\frac{k^2}{2}x^2+4kx+2}}.
\ee
Note that for positive $k$ the denominator remains positive in the 
integration range since
\be
\frac{k^4}{64}x^4-\frac{k^2}{2}x^2+2>0
\ee
as can be checked by solving the corresponding quadratic equation for $x^2$. 
Introducing the new variable $y=k^{1/3}x$ we obtain
\bea
I(k)&\propto&\frac{1}{k^{1/3}} \non \\
& &\times\int_0^{\infty}\!dy\frac{1}
{\sqrt{\frac{k^{8/3}}{64}y^4+4y^3
-\frac{k^{4/3}}{2}y^2
+4k^{2/3}y+2}}.\non \\
& & \label{kinte}
\eea
The integral appearing in (\ref{kinte}) is finite and remains finite for 
 $k=0$. As a result, for small $k$ we have $I(k)\propto k^{-1/3}$
and, therefore,
\be
\frac{1}{\tau_k} \propto k^{5/3}.
\ee
With this $k$ dependence for the relaxation rate the time dependence 
of the correlation function is determined by the following integral
\be
D(t)\propto \int_0^{\pi}dke^{-k^{5/3}Kt}v_k^2. \label{D}
\ee
Here $K$ is a positive constant.
Keeping in mind that $v_k$ is a constant for small $k$ the long time
behavior of $D(t)$ is estimated to be 
\be
D(t)\propto \frac{1}{t^{3/5}}.
\ee
This implies that 
the heat  conductivity coefficient $\kappa$  diverges.
Indeed, we have
\be
\kappa\propto \lim_{t\to\infty}\int_0^{t}d\tau\frac{1}{\tau^{3/5}}
\propto\lim_{t\to\infty}t^{2/5}. \label{divkap}
\ee
Clearly the divergence of $\kappa$  does not mean  that the energy propagates 
through the lattice instantaneously. It just implies that the Fourier heat 
law is not valid  in the infinite FPU $\beta$  lattice.
We can also estimate the dependence of $\kappa$ on the size of the lattice. 
Following reference \cite{Lepri1} we restrict the integral in (\ref{hcc}) to 
times smaller than
 the characteristic time for the energy propagation $N/v_k$. This leads to 
the following $N$ dependence for $\kappa$,
\be
\kappa\propto N^{2/5}. \label{Ndiver}
\ee
Thus we can see that $\kappa$ diverges in the thermodynamic limit which 
is consistent with (\ref{divkap}).

Apart from the $k$ dependence of the decay rate for the mode energy, 
its temperature dependence is also of interest. It follows from 
equation (\ref{pco}) that as a function of temperature the decay rate 
is proportional to $T^2$. 

The estimations we have obtained are in rather good agreement with 
the results of computer simulations.
The divergence of the heat conductivity coefficient
as $N^{\mu}$ with $\mu\approx 0.37$ was observed in the numerical 
studies of the FPU $\beta$ model \cite{Lepri1, Lepri3, Lepri5, Aoki}.
This result is very close to estimation (\ref{Ndiver}). 
Similarly, the $k^{5/3}$ dependence of the decay rate for mode energy was
observed \cite{Lepri3}. The temperature dependence of the decay rate 
reported in \cite{Lepri3} is very close to $T^2$ for weak coupling.

In reference \cite{Lepri3} the same types of $t$ dependence for
the correlation function and $N$ dependence for $\kappa$ were obtained 
by using the mode coupling theory.  
As a function of temperature 
the decay rate obtained through the mode coupling theory behaves as
 $T^{4/3}$ in the limit of weak coupling and as  $T^{1/4}$ for 
strong coupling
\cite{Lepri3}. Thus, the temperature dependence in the weak
coupling limit is different from our  $T^2$ 
estimation. Note, however, that according to Ref. \cite{Lepri3} 
the mode coupling
results should be valid for strong coupling and on very long time scales.
 In general, the mode coupling theory as used in \cite{Lepri3}
allows to make some general statements about the long time behavior of
the current correlation function   for a class of one-dimensional  
lattices while equation (\ref{pco}) gives a more 
detailed picture of the energy equipartition between the normal modes
for the special case of the $\beta$ FPU lattice for the weak coupling 
case. If solved numerically, equation (\ref{pco}) will allow for the
quantitative comparison of the energy equipartition given by the
Peierls equation to the one observed in  computer simulations. 
We will not attempt here to analyze the
relation between our result and the mode coupling theory although this point
clearly deserves attention.

In a recent publication \cite{Narayan} it is claimed that $\kappa$ should 
diverge with system size $L$ as $L^{1/3}$  for all momentum conserving one 
dimensional systems. So far the most careful computer simulation 
\cite{Lepri5} fail to confirm this claim. At present, therefore, this
issue remains unsettled.   

It is well known that for systems such as a gas of hard spheres or Lorentz
gas it is impossible to obtain the correct long time behavior for 
the correlation functions if one uses only the kinetic equation \cite{Balescu1}. This is
because for those systems the spectrum of the collision operator 
is discrete.
As a result if only the kinetic equation is used the long time behavior is
determined by the smallest non zero eigenvalue  of the collision operator and
has an exponential form. In contrast, in our case the collision operator
has a continuous spectrum that is bounded form below by zero. This fact
allows for the non trivial time dependence of the correlation function
to be obtained already in the framework of the kinetic equation.
 
\section{Concluding remarks}
Applying the Peierls equation to the FPU $\beta$ lattice we estimated the 
wave vector and temperature dependence for the decay rate of the 
average mode energy, the long-time behavior 
of the current correlation function and the dependence of the coefficient
of heat conductivity on the size of the lattice. The obtained results 
are in good agreement with the results of the recent 
computer simulations. As we used a number of strong assumptions it can be of 
interest to solve equation (\ref{pco}) numerically in order to verify 
if the assumptions are correct and whether the time evolution of mode 
energies given by (\ref{pco}) is compatible with the results 
of computer simulations for the case of weak coupling.

Recently lattices
with external substrate potentials drew considerable attention
since some of them show finite heat conductivity for $N\to\infty$
\cite{Gillan,Hu1,Hu2,Savin,Lepri4}. We can apply our analysis of
Sec. III to show that the FPU lattice with added  harmonic 
on-site potential of the form  $\sum_ru^2_r$ is likely
to have finite heat conductivity for infinite lattice.
It is easy to show that in this case for $k\to0$ the harmonic frequency 
tends to a constant value  while the group velocity becomes proportional 
to $k$. The energy of the normal mode with $k=0$ is still a constant of 
motion since coefficient (\ref{potential}) vanishes when at least 
one the $k$'s is zero. Therefore, we can expect the decay rate of the mode 
energy behave as $k^\nu$ for small $k$. This will lead to the $t^{-3/\nu}$ 
long time behavior of the current correlation function and, therefore,
finite heat conductivity for $\nu<3$. Thus, if adding the harmonic
on-site potential does not appreciably changes the $k^{5/3}$ wave vector 
dependence of the decay rate, we can expect to find finite heat 
conductivity in this case.
\begin{acknowledgments}
The author would like to thank Dr. Yuriy Pereverzev and Prof. Baowen Li 
for useful comments. Part of this research was supported by grants
from the Robert A. Welch Foundation (Grant No. W-1442) and the Petroleum
Research Fund, administered by the American Chemical Society.
\end{acknowledgments}

\end{document}